\documentclass[aps,prb,a4paper,twocolumn]{revtex4}

\usepackage{graphicx}
\usepackage{dcolumn}
\usepackage{bm}

\begin{document}

\title{Near-field spectroscopy of bimodal size distribution of InAs/AlGaAs quantum dots}

\author{Young-Jun Yu$^{1,2}$}
\author{I. T. Jeong$^{2}$}
\author{J. C. Woo$^{2}$}
\author{Wonho Jhe$^{1,2}$}
\email[Corresponding author:]{whjhe@snu.ac.kr}

\affiliation{$^{1}$Center for Near-field Atom-photon Technology
and $^{2}$School of Physics, Seoul National University, Seoul
151-742, Korea}

\begin{abstract}
We report on high-resolution photoluminescence (PL) spectroscopy
of spatial structure of InAs/AlGaAs quantum dots (QDs) by using a
near-field scanning optical microscope (NSOM). The double-peaked
distribution of PL spectra is clearly observed, which is
associated with the bimodal size distribution of single QDs. In
particular, the size difference of single QDs, represented by the
doublet spectral distribution, can be directly observed by the
NSOM images of PL.
\end{abstract}

\pacs{68.37.Uv, 78.55.Cr, 78.67.Hc}

\maketitle

Self-assembly growth process has raised interesting questions
regarding the growth dynamics of quantum dots (QDs) and the
resulting size and shape distributions.~\cite{aplmo,japliu,woojap}
The size uniformity of InAs/GaAS QDs is typically broken beyond
QD deposition of more than 2.0 monolayer, which is characterized
by the bimodal or multimodal QD
distribution.~\cite{jvcQDsize,aplQDsizebr} In previous works, the
spatial structures of the bimodal QD systems have been
investigated by observing photoluminescence (PL) at various
conditions of QD coverage
growth.~\cite{prbbimodal,prbQDsize,japQDsize,jvcQDsize,japQDsizek}
These results have been valuable for understanding optical
properties of QD ensemble structures grown with various coverage
thickness. However, it is difficult to acquire a high spatial as
well as spectral resolution of single QDs in bimodal QD system if
far-field optical spectroscopy is employed. Several experiments
have been performed by using a near-field scanning optical
microscope (NSOM) for nano-scale optical measurement of
semiconductor QD structures.~\cite{nearapl,saikiprl,yjyuapl,mapl}
Such a unique capability may allow a detailed study of individual
QDs that are grown under various structure conditions. In this
Letter, we report on the observation of high-resolution
near-field laser spectroscopy of bimodal QD size distributions in
the high-density InAs/AlGaAs QD structures.


High density ($\sim$5.5 $\times$10$^{10}$ cm$^{-2}$) InAs QDs,
embedded in 50-nm AlGaAs cladding layers with a 5-nm GaAs capping
layer, were grown by molecular beam epitaxy. The InAs/AlGaAs QD
structure was grown on a 100-nm-thick GaAs buffer layer deposited
on a GaAs (100) substrate.~\cite{woojap} This structure confines
strongly charge carriers within single QDs due to the large
energy difference between InAs QDs and AlGaAs barriers. However,
the size uniformity of QDs is not so good due to the strain
induced by lattice mismatch between InAs QDs and AlGaAs barriers.
Thus the size distribution and deposition thickness of InAs single
QDs is not uniform and depends on the specific location of
nano-scale region on the sample, even though the average coverage
of InAs is about 2.3 monolayer.

For optical pumping of QDs, a Ti:sapphire laser, operated at the
photon energy of 1.67 eV, was used. The excitation laser light
was coupled to a single-mode optical fiber and guided to a
chemically etched sharp fiber tip on which a 100-nm gold-coated
aperture was fabricated.~\cite{jasco} Such a nano-scale light
source, generated by an apertured fiber probe, made it possible to
excite single InAs QDs in small area with nano-scale position
selectivity. The resulting PL signal was collected by the same
fiber so that any loss of spatial resolution due to diffusion
could be minimized.~\cite{eahapl,yjyuapl,matapl} Both the sample
and the apertured fiber probe were enclosed in a He-flow-type
cryostat and kept at 10 K. The PL signal was dispersed by a 0.3 m
single monochromator with a spectral resolution of 0.3 meV and
detected by a liquid-nitrogen-cooled charge coupled device camera
to accomplish the high signal-to-noise ratio.

\begin{figure}[b]
\begin{center}
\scalebox{0.4}{\includegraphics{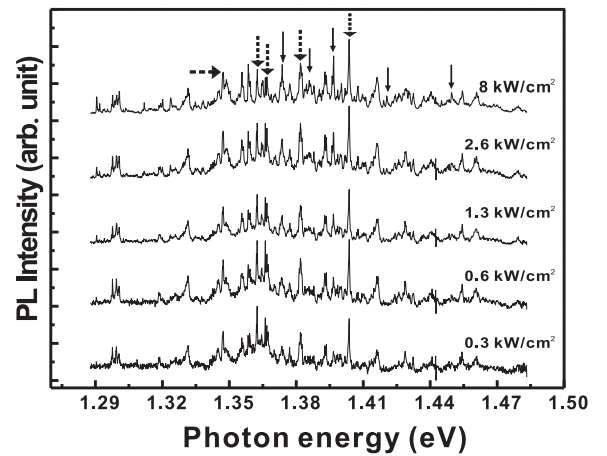}} \caption{Time-integrated PL
spectra collected for 10 s at 10K temperature using an
100-nm-apertured optical fiber tip with increasing the excitation
power density from 0.3 to 8 kW/cm$^{2}$.} \label{fig1}
\end{center}
\end{figure}

Figure~1 shows the time-integrated, high-resolution PL spectra
collected for 10~s at each excitation power density from 0.3 to
8~kW/cm$^{2}$. We have observed many sharp PL peaks of single QDs,
with an average linewidth of about 0.7 meV. It is found that the
PL intensity, resulting from the ground state (dotted arrows), is
saturated as the excitation power density is increased. The PL
from the excited state (solid arrows), on the other hand, emerges
and its intensity increases with excitation power.

\begin{figure}[b]

\scalebox{0.75}{\includegraphics{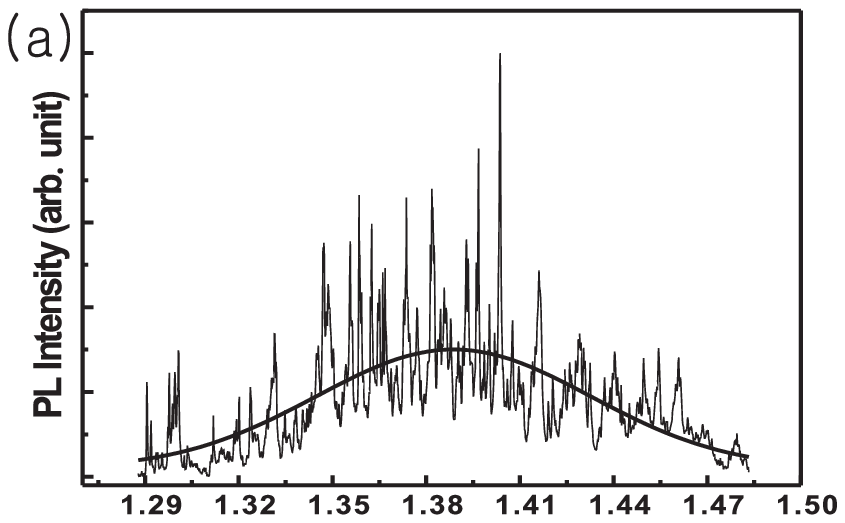}}\label{fig2a}

\scalebox{0.75}{\includegraphics{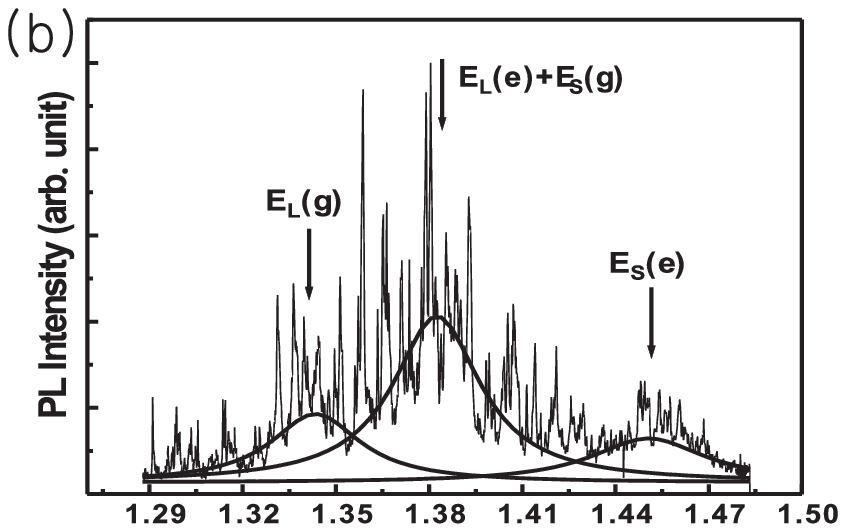}}\label{fig2b}

\scalebox{0.75}{\includegraphics{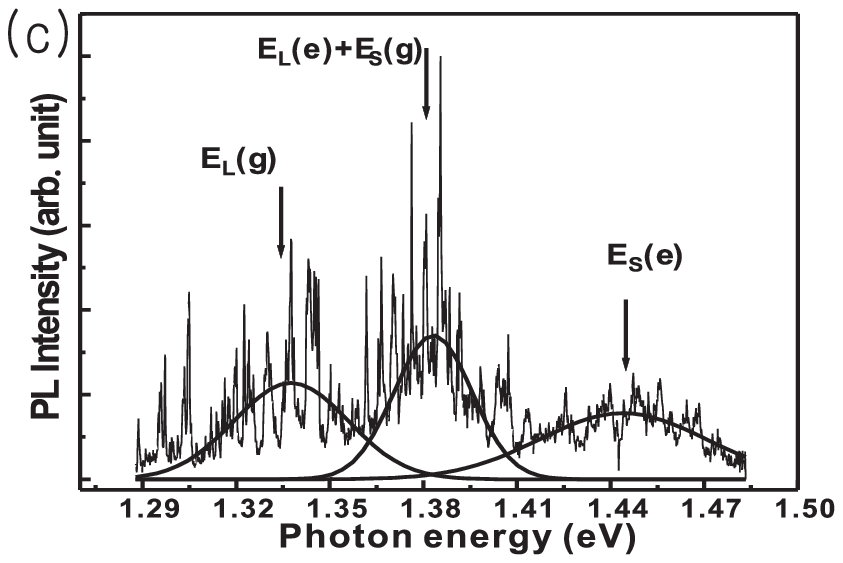}} \caption{Time-integrated
PL spectra on different positions on the same sample collected for
10 s at 10~K using a 100~nm apertured optical fiber tip at the
excitation power density of 8~kW/cm$^{2}$. Where E$_{L}$(g),
E$_{L}$(e), and E$_{S}$(g), E$_{S}$(e) are the ground, excited
states of large size single QDs, and ground, excited states of
small size single QDs, respectively.} \label{fig2c}
\end{figure}

Figure 2 shows the time-integrated PL spectra collected for 10~s,
obtained at different positions on the same sample with
8~kW/cm$^{2}$ excitation power density. In Fig.~2(a), the PL peaks
may be fitted by one Gaussian distribution, which is the same as
that obtained at the excitation power density of 8~kW/cm$^{2}$ in
Fig.~1. However, the PL spectrum in Fig.~2(b), obtained at a
different position, could be decomposed into two Gaussian
distributions (centered at 1.34~eV and 1.38~eV, respectively) and
a shoulder (centered at 1.45~eV). The similar PL peaks in
Fig.~2(c) can be relatively clearly divided into two dominated
Gaussian distributions (centered at 1.33 and 1.38~eV,
respectively) and a shoulder (centered at 1.45 eV). This feature
of Gaussian distributions shown in Figs.~2(b) and (c) are the
characteristic fingerprints of the bimodal size distributions of
single QDs, obtained in far-field PL experiments on the samples
having various QD
thickness.~\cite{aplQDsizebr,aplQDsizeha,prbQDsize,prbbimodal,japQDsize,jvcQDsize,japQDsizek}
Note that the reported energy interval between doublet-like
features of far-field PL spectra due to the bimodal-sized QDs was
about several tens meV,~\cite{japQDsize,aplQDsizebr,prbQDsize} as
similarly observed in our experiment.

Each PL spectrum, which can be fitted into three groups by
Gaussian functions as in Figs.~2(b) and (c), can be attributed to
two contributions of QD size distribution. The first group is the
lowest energy distribution (centered at 1.33~eV in Fig.~2(b) and
1.34~eV in Fig.~2(c)), which is associated with the ground states
of large-sized QDs (E$_{L}$(g)). And the second one, represented
by the middle energy group (centered at 1.38~eV in Fig.~2(b) and
(c)), is contributed by the excited states of large-sized QDs
(E$_{L}$(e)) as well as the ground states of small-sized QDs
(E$_{S}$(g)). The last group is the shoulder group, centered at
1.45~eV in Fig.~2(b) and (c), comes from the excited states of
small-sized QDs (E$_{S}$(e)), as also reported by other
works.~\cite{prbbimodal,jvcQDsize}

We may observe that the thickness of InAs QDs at the sample position
in Fig.~2(b) and (c) is larger than that in Fig.~2(a), because it
was obtained at the lower-energy PL peaks (E$_{L}$(g)) in the
doublet-like PL peak-distribution. The number of large-sized QDs in
Fig.~2(c) is also considered to be larger than that in Fig.~2(b),
because the Gaussian peak at E$_{L}$(g) in Fig.~2(c) is lower by 10
meV and broader by 22 meV than that in Fig.~2(b). Here the center of
energy state and linewidth of E$_{L}$(g) Gaussian peaks are 1.34 eV
and 33 meV [1.33 eV and 55 meV] in Fig.~2(b) [2(c)], respectively.
Thus, we infer that the InAs QDs coverage on position in Fig.~2(c)
is larger than in Fig.~2(b). This is the first time observation of
varying the near-field PL spectra of bimodal size-distributed QDs on
different InAs deposited positions of sample with such a high
spectral resolving power.

\begin{figure}
 \scalebox{0.55}{\includegraphics{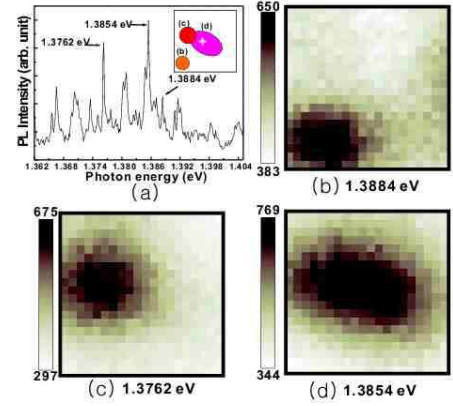}}
\caption{(a) The PL spectrum enlarged spectrum area of Fig.~2(c). PL
images of selective PL peaks of single QDs (b) 1.3884~eV, (c)
1.3762~eV, and (d) 1.3854~eV. The scanning area is
500~nm$\times$500~nm.} \label{fig3}
\end{figure}

The PL spectrum from 1.362~eV to 1.404~eV in Fig.~2(c) is
magnified and shown in Fig.~3(a). This spectrum was collected with
a probe located at the center of scanning PL image area which was
marked by a cross as illustrated in the inset of Fig.~3(a). Note
that there are E$_{L}$(e) peaks and E$_{S}$(g) peaks in this PL
spectrum range (from 1.35~eV to 1.41 eV) in Fig.~2(c). We also
obtain PL images of selected three PL peaks(1.3762~eV, 1.3854~eV,
and 1.3884 eV) at the same area (500~nm$\times$500~nm) in
Fig.~3(b), (c), and (d).

The data acquisition(DAQ) process is as follows:

\noindent (1)The 100~nm apertured probe is located on a selected
sample position and PL spectrum is taken in the entire spectral
range with a nitrogen-cooled CCD camera during 10~s.

\noindent (2)This DAQ process is repeated sequentially in all
other $20\times20$ pixels of the scanning area.

\noindent (3)Then, each PL image at a given energy level from the
400 data files is processed to obtain the corresponding energy
and spatial spectrum of the InAs single QDs.~\cite{eahapl}

The FWHM of horizontal cross section axis on the PL images in
Fig.~3(b), (c), and (d) are about 200~nm, 260~nm, and 400~nm,
respectively. The large-sized single QD with anisotropic PL image in
Fig.~3(d) is due to our growth condition that the average InAs
deposition thickness is about 2.3~ML. Note that if the thickness of
a deposited InAs layer is larger than 2.0 ML, large QDs with a
diameter and height greater than that of the average single QDs size
(in our case, 20~nm lateral dimension and 0.7-nm height) are created
as in the previous observations.~\cite{japliu,jvcQDsize,aplQDsizebr}
We have found that the size of single QD of 1.3854~eV is about twice
larger than that of 1.3884~eV and 1.3762~eV PL peaks in Fig.~3.
Therefore, it seems clear that each selective peaks were divided by
E$_{L}$(e) peaks(centered at 1.3854~eV) and E$_{S}$(g)
peaks(centered at 1.3884~eV, 1.3762~eV) with PL spectrum in
Fig.~2(c) and PL images in Fig.~3. It is very difficult to
discriminate the size difference of single QDs, which are grown in
such a high density ($\sim$5.5~$\times$10$^{10}$~cm$^{-2}$) and
embedded in between barrier and cap layers. Nevertheless, we have
distinguished single QDs with different size having selective PL
energy states from the measured PL images obtained by using an NSOM.

In conclusion, we have observed the variation of high-resolution
PL spectra associated with the size distribution depending on
different locations of QDs structure. We also have characterized
the size-difference of single QDs having PL peaks in doublet-like
distribution by using the PL images obtained by an NSOM.

This work was supported by the Korean Ministry of Science and
Technology through Creative Research Initiatives Program.










\end{document}